\documentclass[fleqn]{mn2e}
\usepackage{graphicx}
\usepackage{amsmath}

\def\msun{{\rm M_{\odot}}}

\def\pb{{P_{\rm b}}}

\title[Blunting the Spike: the CV Minimum Period]
{Blunting the Spike: the CV Minimum Period}
\author[A.R.~King, K. Schenker, J.M. Hameury]{
A.R.~King$^1$, K. Schenker$^1$, J.M. Hameury$^2$\\
1. Theoretical Astrophysics Group, University of Leicester,
Leicester, LE1~7RH, UK\\
2. Observatoire de Strasbourg, UMR 7550 du CNRS, 11 rue de l'Universit\'e,
                67000 Strasbourg, France}

\begin{document}

\maketitle

\begin{abstract}
The standard picture of CV secular evolution predicts a spike in the
CV distribution near the observed short--period cutoff $P_0 \simeq
78$~min, which is not observed. We show that an intrinsic spread in
minimum (`bounce') periods $P_{\rm b}$ resulting from a genuine
difference in some parameter controlling the evolution can remove the
spike without smearing the sharpness of the cutoff. 
The most probable second parameter is different
admixtures of magnetic stellar wind braking (at up to 5
times the GR rate) in a small tail of systems, perhaps implying that
the donor magnetic field strength at formation is a second
parameter specifying CV evolution. We suggest that
magnetic braking resumes below the gap with a wide range, being well
below the GR rate in most CVs, but significantly above it in a small
tail.
\end{abstract}



\section{Introduction}

The orbital period $P$ is the one parameter of a cataclysmic variable
(CV) which observers can usually measure with confidence. Roche
geometry implies a close relation
\begin{equation}
P \propto \biggl(\frac{R_2^3}{M_2}\biggr)^{1/2}
\label{porb}
\end{equation}
between this period and the mass and radius $M_2, R_2$ of the
mass--losing star. The accretion luminosity of CVs suggests that the
mass transfer timescale $t_{\rm M} = -M_2/\dot M_2$ is considerably
shorter than the age of the Galaxy, so that $M_2, P$ change on this
timescale. Hence the observed distribution of CV periods is by far the
most significant indicator of CV evolution. As is well known, it is
consistent with the idea that the evolution is driven by angular
momentum losses from the binary orbit (see e.g. King, 1988 for a review).

The observed CV histogram (Fig. 1) cuts off sharply at an orbital
period of $P = P_0 \simeq 78$~min. There are respectively 0 and 12
systems in the period ranges $P_0 \pm 5$~min. The idea that $P_0$
represents a global period minimum ($\dot P = 0$ for $P_0$) for CVs
has been widely accepted for the last two decades. It is clear that
such a global minimum can exist (Paczy\'nski, 1981; Paczy\'nski \&
Sienkiewicz, 1981; Rappaport, Joss \& Webbink, 1982). As the mass
$M_2$ of an unevolved secondary star in a CV is reduced by mass
transfer, the binary period $P$ usually decreases also, with $R_2 \sim
M_2 \sim P$. However for very small $M_2 \la 0.1\msun$, the
secondary's Kelvin--Helmholtz time $t_{\rm KH}$ exceeds the timescale
$t_{\rm M} = -M_2/\dot M_2$ for mass transfer driven by gravitational
radiation. Instead of shrinking smoothly to the main--sequence radius
appropriate to its reduced mass, the star cannot now reduce its
entropy quickly enough and contracts more slowly. From (\ref{porb}) we
see that once $R_2$ decreases more slowly than $M_2^{1/3}$ the orbital
period $P$ must begin to increase, defining a minimum (`bounce')
period $P_{\rm b}$ for the system. It is important to realize that
there is nothing extreme about conditions in the binary at this point;
in particular the mass transfer rate remains almost precisely the same
($\sim 4\times 10^{-11}\msun~{\rm yr}^{-1}$) as when the system was
well above the period minimum, cf Kolb (2002), Fig. 3. Nor do
conditions change drastically thereafter: since $M_2 \simeq 0.1\msun$
at $P_0$ we have $t_{\rm M} \sim 2.5$~Gyr. Hence most CVs do not have
time to evolve to significantly longer periods or lower mass transfer
rates after reaching $\pb$: clearly to evolve to $-\dot M_2 =
1\times 10^{-11}\msun~{\rm yr}^{-1}$ would require of order the age of
the Galaxy. CVs reaching $P_0$ thus remain clustered there with mass
transfer rates similar to those of systems at longer periods.

Detailed calculations always predict a value $P_{\rm b}$ very close
to, if slightly shorter than, the observed $P_0$. The discrepancy
$P_{\rm b} < P_0$ is persistent, but may reflect uncertain or
over--simple input physics (cf Kolb \& Baraffe, 1999). In particular
the difference between the true and spherically approximated radii
may account for most of the disagreement. However there is a much more
serious problem with this interpretation of the observed cutoff at
$P_0$.  This concerns the discovery probability
\begin{equation}
p(P) \propto \frac{(-\dot M_2)^{\alpha}}{|\dot P|}.
\label{prob}
\end{equation}
Here $\alpha$ is some (presumably positive) power describing
observational selection effects. For example $\alpha = 3/2$ for a
bolometric flux--limited sample, while $\alpha = 1$ is often
assumed for systems detected via optical outbursts. Since $\dot P = 0$ at
$P = P_{\rm b}$, $p(P)$ must clearly have a significant maximum there
unless $-\dot M_2$ declines very sharply near this period. In other
words, the observed CV period histogram should show a sharp rise near
a global minimum $P_{\rm b}$ unless the mass transfer rate drops
there. However, as mentioned above, all evolutionary calculations show
that $-\dot M_2$ changes very little as $P_{\rm b}$ is approached. We
conclude that if $P_0$ is a global minimum there should be a large
`spike' in the CV period histogram there (cf Kolb \& Baraffe,
1999). Instead, the observed period histogram (Fig. 1) has a `corner'
at $P_0$, i.e. a sharp cutoff of a fairly flat distribution, but no
spike.

\begin{figure}
\centerline{\includegraphics[clip,width=0.85\linewidth]{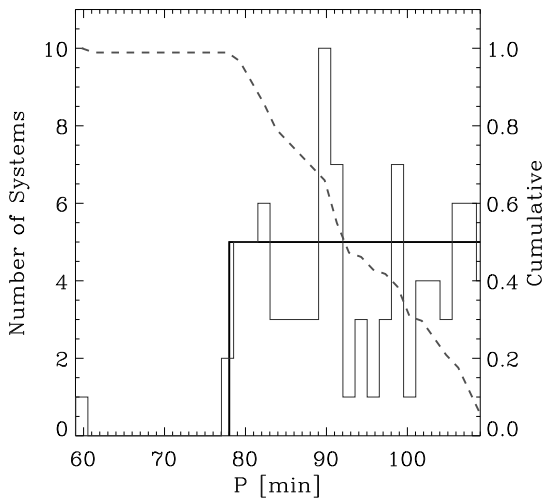}}
\caption{CV period histogram and cumulative distribution (dashed)
below the observed period gap (data from Ritter \& Kolb, 1998).  The
Heaviside function shown is the representation (\ref{hist}).}
\end{figure}

The lack of a spike in the observed distribution has prompted numerous
theoretical investigations (see Kolb, 2002 for a review, and Barker \&
Kolb 2002). Most of these propose ways in which CVs might become
difficult to discover near $P_0$. A basic problem for this type of
argument is that, as we have seen, there is nothing at all unusual
about the system parameters (mass transfer rate, separation etc) at
this period. Further, attempts to use accretion disc properties as a
way of making systems hard to discover founder on the fact that the
AM~Herculis systems, which have no accretion discs, have precisely the
same observed short--period cutoff $P_0 \simeq 78$~min, and no spike
either. King \& Schenker (2002) suggested that CV formation may take
roughly the age of the Galaxy, so that the oldest systems have not yet
quite reached the minimum period. The fine--tuning here is perhaps
worrying, although not easily disproved.
In this paper we 
suggest a solution to the spike problem invoking a
a different new ingredient in the standard picture of CV evolution.

\section{Blunting the Spike}

\begin{figure}
\centerline{\includegraphics[clip,width=0.82\linewidth]{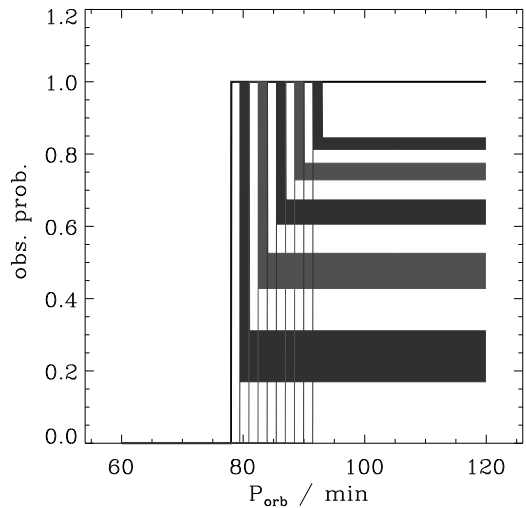}}
\caption{Simple construction of a square distribution by adding
  discrete L-shaped individual components. The parameters used for the
  case shown are: $r = 10, a = 0.83, \Delta P = 1.5, P_0 = 78$.
}
\end{figure}
The new element we introduce into the standard picture is the idea of
a {\it distribution} of minimum periods $\pb$. At first sight this may
not seem a particularly radical step. However it amounts to allowing
the intervention of a second controlling parameter.
Second, note that the difficult part here is not smearing out the
spike, but {\it simultaneously retaining the sharp cutoff at $P_0$.}

To fix ideas we first formulate a simplified version of the
spike problem which has the virtue of being exactly soluble, and then
proceed to a more realistic approach. In the simplified problem we
make the following approximations:

1. We assume that the relative discovery probability for any individual CV
   depends only on $P - \pb$, and can be represented as
\begin{equation}
p(P - \pb) = H(P - \pb) - aH(P - \pb - \Delta P). 
\label{pp}
\end{equation}
Here the $H$'s are Heaviside functions, $\pb$ is the bounce period for
the individual CV, and $a, \Delta P$ are constants. As can be seen by
comparison with the lowest panel of Figure 3 of Kolb (2002) this is a
fair representation if we take $a \simeq 0.83, \Delta P \simeq
1.5$~min, with $\pb = 67$~min in the case shown there. 

2. The observed CV histogram is taken as
\begin{equation}
N(P) = N_0H(P - P_0)
\label{hist}
\end{equation}
where $N_0$ is a normalization. This is a relatively crude
representation of the observed histogram (Fig. 1), but does capture in
essence its only significant feature, the `corner' at $P_0$.

Armed with these assumptions, we determine the relative distribution
$n(\pb)$ of CVs with minimum period at a given $\pb$. Clearly the
observed histogram $N(P)$ is the convolution of $n(\pb)$ with $p(P -
\pb)$, so using (\ref{pp}, \ref{hist}) we have
\begin{equation}
N(P) = \int_0^P{\rm d}\pb n(\pb)p(P - \pb) = N_0H(P - P_0).
\label{int}
\end{equation}
Taking Laplace transforms, the convolution theorem gives
\begin{equation}
\bar n(s) \bar p(s) = \frac{N_0}{s}e^{-sP_0}.
\end{equation}
Now since 
\begin{eqnarray}
\lefteqn{\bar p(s)  = \int_0^{\infty}e^{-sx}[H(x) - aH(x - \Delta
 P)]{\rm d}x} \nonumber \\
& &  = \frac{1}{s} - \frac{a}{s}e^{-\Delta P s}
\end{eqnarray}
we get 
\begin{equation}
\bar n(s) = \frac{N_0e^{-sP_0}}{1 - ae^{-s\Delta P}}
= N_0\sum_{r=0}^{\infty}a^re^{-(P_0 + r\Delta P)s}.
\end{equation}
Hence transforming back we have
\begin{equation}
n(\pb) = N_0\sum_{r=0}^{\infty}a^r\delta(\pb - P_0 - r\Delta P).
\label{sum}
\end{equation}
This result gives a decomposition of the assumed histogram which is
completely obvious when plotted (Fig. 2). However it does illustrate
the main point: one can create a rather flat but cut--off distribution
(cf \ref{hist}) from the very spiky individual probabilities
(\ref{pp}) by taking a distribution of bounce periods $\pb > P_0$
falling off with $\pb$ (here as $a^{(\pb/\Delta P)}$) for $\pb >
P_0$. Dropping the higher--order terms $r \geq r_{\rm max}$ produces a
downward step at $P = P_0 + r_{\rm max}\Delta P$ in an otherwise
uniform distribution, the discontinuity having size $a^{r_{\rm max}}$. 
If a 15\% downwards glitch is regarded at the limit of
acceptability when compared with observation, we can estimate the
spread in $\pb$ required to wash out the spike as $r_{\rm max}\Delta P
= (\log 0.15/\log 0.83)\times 1.5$~min = 15~min. Similarly if we
define $r_{1/2}$ by $a^{r_{1/2}} = 0.5$ then more than 50\% of CVs in
the period range $P_0 < P < P_0 + r_{1/2}\Delta P$ are close to their
local minima $\pb$ (i.e. are in their local `spikes'). We find
$r_{1/2} = 3.72$, so we expect half of the CVs in a 5.6~min range
above $P_0$ to be near their local $\pb$ values.

\begin{figure}
\centerline{\includegraphics[clip,angle=-90,width=0.95\linewidth]{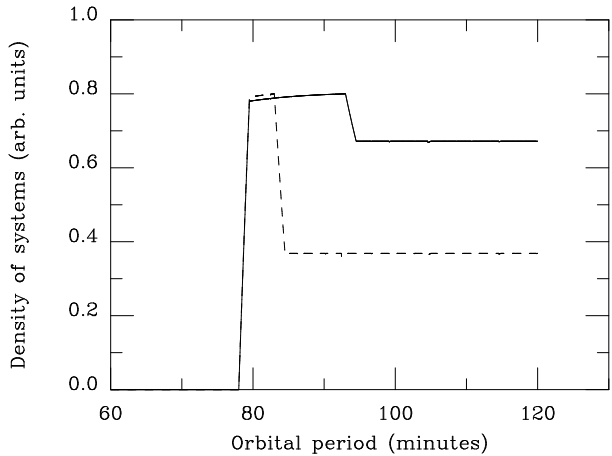}}
\caption{Predicted distribution, assuming (a) that $\pb$ varies over a 
  15 min range, starting from 78 min, with $n(\pb)$ proportional to
  $\exp(-k \, (\pb -P_0) / \Delta P$, $k=1.8$, and (b) $\pb$ varies over a 
  5 min range, with $k=0.6$. In both cases, the intrinsic width of the
  spike is 1.5 min.
}
\end{figure}

It is now easy to guess a continuous form of $n(\pb)$ which gives a
reasonable representation of the corner
as $n(\pb) = \exp[-0.124(\pb - P_0)]$.
The result is shown in Fig. 3, with the
fitting extended to a maximum $\pb$ equal to (a) $P_0 + 5~{\rm min}$,
and (b) $P_0 + 15~{\rm min}$. As can be seen, extending to the longer
period produces a much more acceptable representation.

We can thus state our main result: the CV distribution
near the short--period cutoff $P_0$ is well reproduced if CVs have a
declining distribution of individual minimum periods $\pb$ above the
floor value $P_0$. 
\section{The hidden parameter}

In standard CV evolution (e.g. King, 1988) the global properties of a
system at any epoch are specified by two parameters, e.g. the initial
primary and secondary masses $M_{1i}, M_{2i}$, corresponding to a
white dwarf and an unevolved low--mass star respectively. Given these
two quantities we can in principle calculate all others such as
period, mean mass transfer rate and discovery probability $P, -\dot
M_2, p(P)$ at any subsequent time. In practice evolutions starting
from different initial secondary masses $M_{2i}$ converge very rapidly
to an effectively common track (Stehle et al., 1996), and even the
dependence on white dwarf mass is very weak. Thus in this standard
picture all global properties of a CV are essentially specified if we
know the orbital period $P$. This extreme simplicity gives the
standard picture its predictive power, but also makes it vulnerable to
problems such as the missing spike at $P_0$. This is clearly predicted
if every CV follows the common evolutionary track. To blunt the spike
in the manner considered above requires sensitivity to a further
physical parameter. 
Paczy\'nski \& Sienkiewicz (1983) show that this cannot
be the primary mass or the assumed mean surface opacities 
(in any case the latter do not in
principle constitute an independent parameter).

We therefore turn to the other effect altering $\pb$ considered by
Paczy\'nski \& Sienkiewicz (1983), namely increasing systemic angular
momentum losses $\dot J$ by a factor $f$ above the gravitational
radiation value $\dot J_{\rm GR}$. They found a $\sim 10$~min
spread in $\pb$ for a range $1 \leq f \la 2$. We checked this using
the codes of Mazzitelli
(1989) -- as adapted by Kolb \& Ritter (1992) -- and Hameury (1991),
and found that the required $\ga
15$~min spread results for $1 \leq f \la 6$. 

Note that in fixing this range of $f$ we assume that evolution with
$f=1$, i.e. driven purely by GR, does produce the minimum bounce
period $\pb = P_0 \simeq 78$~min, and thus that theoretical efforts in
this direction will succeed. We could instead in principle use a value
$f_{\rm min} > 1$ to bring $\pb$ and $P_0$ into agreement, but this
would rob us of the natural explanation for such a global minimum
$f_{\rm min}$, namely that this is the limit in which all other
angular momentum loss mechanisms are small compared with GR.

\section{Discussion}

We have shown that the corner in the CV period distribution is
reasonably well reproduced provided that CV evolution has a second
controlling parameter spreading the minimum periods of an exponential
tail of CVs by $\ga 15$~min. The most likely candidate for this
parameter is an increase of the orbital angular momentum loss rate
above that provided by gravitational radiation by a factor $f$,
ranging from 1 to at least 6. Since gravitational radiation must
always be present, the requirement $f \geq 1$ simply implies another
angular momentum loss process whose strength ranges from values $<<
|\dot J_{\rm GR}|$ to $\ga 5|\dot J_{\rm GR}|$. An obvious candidate
here is varying admixtures of magnetic stellar wind braking. The
physical origin of the extra degree of freedom specified by $f$ might
then simply be the intrinsic strength of the stellar magnetic field in
the region of the ISM from which the secondary star formed, which may
then influence the strength of any dynamo--amplified or generated
field. It is an observed fact that otherwise similar stars can have
very different magnetic field strengths, and this presumably applies
to the secondary stars of CVs also. Apparently in most cases this
field gives only weak magnetic braking $|\dot J_{\rm MB}| << |\dot
J_{\rm GR}|$, but there is evidently a tail of systems with stronger
braking. This would for example naturally result if the distribution
of intrinsic magnetic fields were a gaussian, with the mean
sufficiently high to give strong braking above the period gap, but too
low for such braking below the gap in all but an exponential tail of
CVs.

\begin{figure}
\centerline{\includegraphics[clip,width=0.95\linewidth]{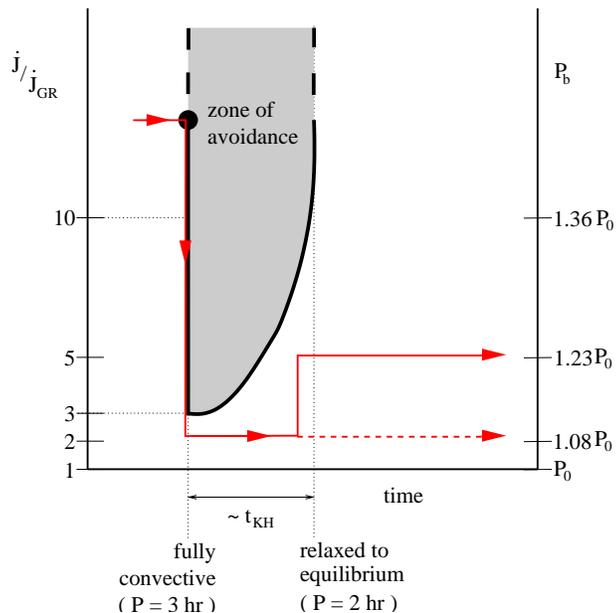}}
\caption{Forbidden zone (shaded area) for angular momentum losses
$\dot{J}$ in the interrupted magnetic braking picture (Rappaport,
Verbunt \& Joss, 1983, Spruit \& Ritter, 1983) illustrated
schematically for a CV becoming fully convective at $P \simeq 3$~hr
with magnetic braking rate $\dot J_0 \gg 10 \, \dot{J}_{\rm GR}$. To
form the observed period gap between $\sim 3$ and $\sim 2$ hours,
$\dot{J}$ must drop rapidly from a value of $\dot{J}_0$ to $\la 3 \,
\dot{J}_{\rm GR}$, and {\em stay low} for a Kelvin time. After this,
any amount of magnetic braking may be possible, leading to
significantly different bounce periods $\pb$ as given on the righthand
axis.
}
\end{figure}

\noindent
This is a satisfying conclusion, as one of the objections to the usual
mechanism for forming the CV period gap between 3 -- 2 hours has been
that strong magnetic activity is observed in main--sequence stars with
masses low enough to be secondaries in CVs below the gap. The answer
to this objection is now that indeed such activity may well occur in
systems below the gap, but in most cases the resulting braking is
weaker than GR. However for $f \sim 6$ much
of the tail has values of magnetic braking considerably stronger than
the extrapolation of the usual magnetic braking laws adopted for CVs
above the gap (Verbunt \& Zwaan, 1981 Mestel \& Spruit,
1987). Evidently these laws do not describe this regime well, in
agreement with the idea that the nature of the braking changes
character once the secondaries become fully convective (Taam \&
Spruit, 1989). It is therefore also plausible that intrinsically
stronger fields are required below the period gap in order to give
significant braking rates, thus accounting for the exponential tail of
systems with strong braking as suggested above.

We note that gap formation requires only that magnetic braking $\dot
J_{\rm MB}$ should be {\it interrupted}, not {\it suppressed},
at $P \simeq 3$~hr (plausibly where the secondary first becomes fully
convective).
Figure 4 illustrates cases where the braking
after the detached phase almost reaches the value before it.
Provided magnetic braking does not resume, at something like its
former strength, before the oversized secondary has managed to shrink
significantly towards its main--sequence radius (on a timescale $\sim
t_{\rm KH} \sim 10^8$~yr), a gap of the observed properties will
result. If the association with the change to a fully convective
structure is maintained, we would thus require that the resulting
change in field topology or strength takes at least a time $\sim
t_{\rm KH}$.
To achieve the required smearing in $\pb$ the angular momentum losses
of a fraction of CVs must change in a way similar to the curve shown
in Fig.~4.

This discussion suggests that there is no obvious reason for any upper
limit on the factor $f$, given only that the distribution of $f$
values drops as $f$ increases above unity. Systems with large $f$ have
high mass transfer rates, and bounce at quite long periods. This
is perfectly consistent with observation, provided that only a
minority of CVs below the gap have such high values of $|\dot J_{\rm
MB}|$.  We thus suggest that magnetic braking has the following
properties (cf Fig.~4)

(i) in most CVs above the period gap it drives mass transfer rates
$-\dot M_2 \ga 5\times 10^{-9}\msun~{\rm yr}^{-1}$

(ii) when a CV secondary becomes fully convective, magnetic braking
rapidly drops to very low values for at least the thermal timescale of
the star

(iii) magnetic braking is present in all CVs below the period gap, and
its strength depends on an inherent property of the secondary,
possibly the formation magnetic field. In most cases it is
weaker than gravitational radiation, but there is an exponential tail
of systems with much stronger braking. These values are much bigger
than given by extrapolating the usual magnetic braking laws to short
orbital periods, and may thus require the strongest 
fields.

\section{Acknowledgments} 

Theoretical astrophysics research at Leicester is supported by a PPARC
rolling grant. ARK thanks the Observatoire de Strasbourg for
hospitality.

\end{document}